\documentclass[aps,prl,reprint,superscriptaddress,nobibnotes]{revtex4-1}
\usepackage{amsmath}
\usepackage{bbm}
\usepackage{commath}
\usepackage{graphicx}
\usepackage{color}
\usepackage{indentfirst}
\usepackage{setspace}
\usepackage{tabularx}
\usepackage{multirow}
\usepackage{subfigure}

\begin{document}

\title{A Tetrahedron-tiling Method for Crystal Structure Prediction} 

\author{Qi-Jun Hong}
\email[e-mail:]{ qhong@alumni.caltech.edu}
\affiliation{School of Engineering, Brown University, Providence, Rhode Island 02912, USA}

\author{Joseph Yasi}
\affiliation{The MathWorks, Inc., 3 Apple Hill Drive,
Natick, Massachusetts 01760, USA}

\author{Axel van de Walle}
\affiliation{School of Engineering, Brown University, Providence, Rhode Island 02912, USA}
\date{\today}

\begin{abstract}
Reliable and robust methods of predicting the crystal structure of a compound, based only on its chemical composition, is crucial to the study of materials and their applications.
Despite considerable ongoing research efforts, crystal structure prediction remains a challenging problem that demands large computational resources.
Here we propose an efficient approach for first-principles crystal structure prediction. The new method explores and finds crystal structures by tiling together elementary tetrahedra that are energetically favorable and geometrically matching each other. This approach has three distinguishing features: a favorable building unit, an efficient calculation of local energy, and a stochastic Monte Carlo simulation of crystal growth. 
By applying the method to the crystal structure prediction of various materials, we demonstrate its validity and potential as a promising alternative to current methods. 
\end{abstract}
\maketitle

\singlespacing

Crystal structure prediction (CSP) \cite{Mad88, Desiraju02,Woodley08} has been a topic of great interest with 
both fundamental importance in physical science -- an ambitious goal to predict the structure of a crystal from the knowledge of its chemical composition --
and enormous potential in modern materials design and discovery ranging from petrochemical industry \cite{Norskov09}
and electrical energy storage \cite{Wolverton12} to the study of phase diagrams \cite{vandeWalle02}.
This recognition has led to the development of a number of ingenious computational approaches, including 
simulated annealing \cite{Kirkpatrick83,Cerny85}, basin hopping \cite{Wales99}, evolutionary algorithms \cite{Coley99,Oganov06,Zunger08,Oganov13,Calypso}, high-throughput data mining \cite{Ceder03,Liu04}, \textit{etc}.
Over the past decade, contemporary CSP methods started to develop genuine predictive capability and they keep improving.
However, structure prediction remains a challenging problem that demands large computational resources. 
To address this issue, we propose an alternative CSP approach, by which new crystal structures are predicted by tiling from energetically favorable and geometrically matching elementary tetrahedra.
This strategy proves advantageous because (i) tetrahedra energies can be efficiently determined from a relatively small number of quantum mechanical calculations and (ii) tetrahedra form natural building blocks for arbitrary crystal structures.

\begin{figure}[]
\centering
\includegraphics[width=0.35\textwidth]{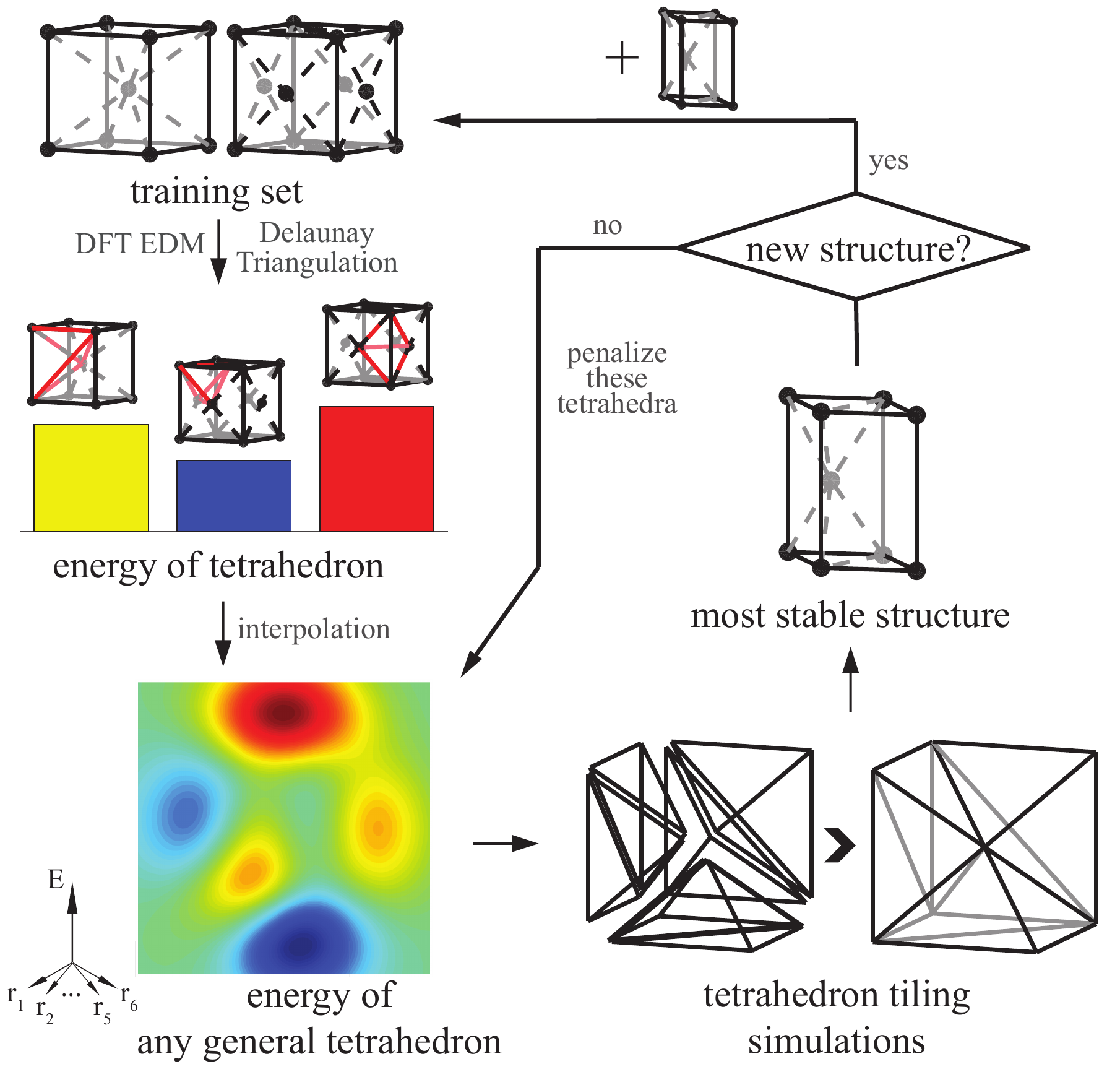}\\
\caption{\label{diagram}
A flowchart showing the mechanism of crystal structure prediction we propose.}
\end{figure}

We propose a tetrahedron-tiling method which we illustrate with the flowchart shown in Fig. \ref{diagram}.
Starting from a small training set of common crystal structures, we first calculate the local energies of the tetrahedra from these structures, based on local energy density method (EDM) \cite{Martin92, Trinkle11} in density functional theory (DFT) \cite{Kohn64,Kohn65,Jones89} and the Delaunay triangulation \cite{Del34}.
These data points of known tetrahedra enable us to interpolate, at a very low cost, an empirical energy landscape for any tetrahedron with general coordinates.
Based on this tetrahedron energy landscape, we run Monte Carlo (MC) simulations to tile together tetrahedra that are favorable in energy and match each other in geometry.
The simulations generate a series of stable crystal structures, which are further added to the training set to provide more data points on new tetrahedra, systematically improving the empirical tetrahedron energy landscape.
After sufficiently many cycles, CSP is achieved since all important tetrahedra, hence crystal structures as well, are captured and included in the training set.
The set of candidate crystal structures generated in this fashion is then further screened via DFT, to yield the final predicted low energy structure(s)
with full DFT accuracy.

This approach offers two distinct advantages.

First, slicing crossover in existing evolutionary algorithms (e.g., Ref. \onlinecite{Oganov06}) inevitably involves undesirable lattice and structure mismatches. 
When slicing crossover combines two parental structures with high fitness ranking to create offsprings, one key idea is to put together two locally favorable structures to increase the chances of producing a better chemistry.
Indeed, structural formation energies are typically dominated by short-range interactions, and this wisdom is widely adopted in CSP methods. 
Despite being an ingenious idea, slicing crossover sometimes creates mismatches of lattice and chemical bonds, and hence the slicing plane may become an undesirable interface with an associated energy penalty.
We note that there is a better way to combine local structures naturally:
a deformable tetrahedron is an ideal tiling unit.
Geometrically, a tetrahedron is the simplest and most elementary form of a polyhedron in three-dimensional space. Any set of three-dimensional points in general position can be uniquely decomposed into tetrahedra by the Delaunay triangulation \cite{Del34}.
Energetically, a tetrahedron has four vertices, so it accounts for local four-body interactions, which naturally incorporates key chemical parameters, such as bond length, bond angle, dihedral angle, solid angle, \textit{etc}.
Furthermore, recent development of local EDM \cite{Martin92, Trinkle11} in DFT enables us to compute a local effective gauge-invariant energy for each atom in a crystal and assign a well-defined energy to each tetrahedron unit via a geometric construction without undergoing a large-scale fitting exercise typically associated with the construction of effective interatomic potential energy model.
These capabilities open up a new avenue to CSP: it becomes possible to build a structure by tiling together tetrahedra that are locally favorable and geometrically match with each other.

Second, contemporary CSP approaches are not sufficiently efficient in the context of first-principles calculations, and this becomes an obstacle to their application in high-throughput materials screening and discovery.
Historically, the realization that a brute-force systematic search through all possible structures is infeasible due to the exponential complexity of the optimization problem leads to the wide adoption of more pragmatic techniques such as evolutionary algorithm and simulated annealing.
While this strategy systematically improves the expected search time and increases the probability of finding the global minimum,
a trial and error approach with an energy model as expensive as DFT is by no means cost-effective.
For instance, evolutionary algorithms from fully first-principles demand a large amount of energy evaluations, which renders it prohibitively expensive for complex structures \cite{USPEX, Calypso}.
For simulated annealing, running \textit{ab initio} molecular dynamics (MD) or MC is already a heavy task, not to mention the requirement of slow cooling and hence a long trajectory.
By contrast, we devise an approach based on an effective energy model of tetrahedron, which requires only a small set of DFT calculations performed before hand.
This property makes it highly efficient in the exploration of possible tetrahedron tiling.

An important aspect of the method is the mapping of a solid structure to a set of tetrahedra.
In principle, a solid structure $\textbf{R}^m$ with general coordinates ($\textbf{R}^m$ is a $3\times m$ matrix of atomic coordinates, $m$ is the number of atoms) can be uniquely decomposed into tetrahedra through the Delaunay triangulation \cite{Del34}.
The tetrahedra serve as elementary building blocks which comprise the solid structure as a whole.
After the Delaunay triangulation, the solid structure $\textbf{R}^m$ is represented by $n$ tetrahedra $\mathbbm{r}=$\{$\textbf{r}_1,\textbf{r}_2,\cdots,\textbf{r}_i,\cdots,\textbf{r}_n$\}.
We therefore write the total energy of the solid $\textbf{R}^m$ as a sum of the local energies of tetrahedra $\mathbbm{r}$,
\begin{eqnarray}
\textbf{R}^m &\rightarrow& \mathbbm{r}=\{\textbf{r}_1,\textbf{r}_2,\cdots,\textbf{r}_i,\cdots,\textbf{r}_n\}, \label{eq_decomp}
\\
E_{\text{tot}} \left( \textbf{R}^m \right) &\approx& \sum_{i=1}^{n}{E(\textbf{r}_i)} \text{, } E(\textbf{r}_i) = {\epsilon(\textbf{r}_i) \Omega_s(\textbf{r}_i)}/4\pi,\label{eq_add}
\end{eqnarray}
where ${E(\textbf{r}_i)}$ is the local energy of a tetrahedron $\textbf{r}_i$, $\epsilon(\textbf{r}_i)$ is the corresponding specific energy (in eV/atom), and $\Omega_s(\textbf{r}_i)$ is the sum of its four solid angles, as $\Omega_s(\textbf{r}_i)/4\pi$ accounts for the fractional number of atoms within a tetrahedron.
The tetrahedron energy is equal to the sum of atom-specific energies found via EDM, each weighted by the solid angle that expresses the fraction of each atom actually lying inside the tetrahedron.
For multicomponent systems, composition is controlled by changing chemical potentials of elements.
We add a term of the form $-\mu^T x$ to the Hamiltonian (where $\mu$ and $x$ are, respectively, the vectors of chemical potentials and compositions) and work with a grandcanonical ensemble (\textit{i.e.}, possible moves in our MC simulations include atom deletion or atomic species changes).

Converting a bulk solid structure into a set of \textit{independent} tetrahedra greatly simplifies the problem.
While the former has $3m$ degrees of freedom (or $3m-6$, if excluding translation and rotation), each tetrahedron of the latter has only six, drastically reducing the complexity.
In particular, it becomes feasible to explore the full energy landscape of tetrahedron $\epsilon(\textbf{r})$, given the low dimensions.

As a key component of the method, constructing this energy landscape enables us to evaluate the specific energy $\epsilon(\textbf{r}_i)$ of any general tetrahedron $\textbf{r}_i$.
This is achieved in two steps, (1) training, \textit{i.e.}, rigorous DFT-EDM calculations of $\epsilon$ for a group of common structures (which form a training set), and (2) interpolation for a general tetrahedron based on the known tetrahedra obtained in the training step, i.e.,
\begin{equation}\label{eq_interpol}
\epsilon(\textbf{r}) = \sum_{k=1}^N{ \epsilon(\textbf{r}_k^0) w(\textbf{r}-\textbf{r}_k^0)} \bigg/ \sum_{k=1}^N{w(\textbf{r}-\textbf{r}_k^0)} ,
\end{equation}
where $w$ is a weighting function (that takes its maximum value at $|\textbf{r}-\textbf{r}_k^0|=0$ and decays smoothly as $|\textbf{r}-\textbf{r}_k^0|\rightarrow \infty$), $\textbf{r}$ is a general tetrahedron, $\textbf{r}_k^0$ is a known tetrahedron, and $N$ is the total number of known tetrahedra in the training set (see Supplemental Material).

After building a full energy landscape of tetrahedra $\epsilon(\textbf{r})$, we can estimate energy of any structure base on $\epsilon(\textbf{r})$. In detail, an unknown structure is first decomposed into elementary tetrahedra, whose specific energies are evaluated from $\epsilon(\textbf{r})$. The energies of individual tetrahedra add up to the total energy of the structure according to Eqs. (\ref{eq_decomp}) and (\ref{eq_add}) .
This capability is employed to simulate tetrahedra tiling and crystal growth (see Supplemental Material), with an ultimate aim to achieve CSP.
We note that our tetrahedra energies represent energies in bulk material. 
So even though the growing cluster has a surface, there is actually no surface energy penalty associated with it. 
This is a desirable property, because the surface will not bias the search for the most stable bulk structure. 

The tetrahedron energy landscape effectively serves as an empirical potential in tetrahedron tiling simulations.
Tetrahedra contain crucial geometric information such as bond length, bond angle, \textit{etc}.
Indeed, they take account of four-body interactions among first nearest neighbors, as each tetrahedron connects a set of four neighbors. 
Hence these tetrahedra in principle dominate energetic properties of the solid structure.
Replacing total energy of a structure by a sum of tetrahedra's local energies, to be rigorous, introduces an approximation, as this process attempts to capture real many-body interactions through local energies derived from an empirical tetrahedron energy landscape.
Only interaction among the first few nearest neighbors are taken into account and long range interactions are truncated.
The real total energy will inevitably differ from the sum.
For instance, the mapping from a structure $\textbf{R}^m$ to its tetrahedron set $\mathbbm{r}$ is not injective, \textit{i.e.}, two structures may correspond to the same set of tetrahedra, and hence two identical tetrahedra could have different local energies in different structures.
Nevertheless, we note that our interpolation scheme actually handles this by smoothing: this tetrahedron is then assigned the average value for all identical tetrahedra.
Furthermore, the final relative stability will be determined by DFT, while the empirical energy landscape $\epsilon(\textbf{r})$ is employed only to narrow the search down to a few candidate stable structures.
While this approach of tetrahedron energy density evaluation appears similar to interatomic potential construction and fitting,
the latter process,  to achieve comparable accuracy of four-body interaction, would typically require (1) specifically developed analytic functional forms and (2) an enormous number of structures in the training set, which both demand heavy human input and computer resources.
The EDM approach hence far outperforms it in terms of simplicity and computer cost.


We note that there are two remaining hurdles and we address them using robust strategies.

The first problem lies in the process of tetrahedron tiling.
Since the tetrahedron energy landscape $\epsilon (\textbf{r})$ is an approximation to DFT interactions,
the structure built by tetrahedron tiling is not necessarily the most stable one in the DFT aspect.
Furthermore, the simulation may undesirably generate a known structure which is already in the training set, thus achieving little more.
This problem is avoided via a simple but powerful exploration algorithm: we ask the simulation to produce a series of
distinct stable structures, and let DFT decide their final relative stability.
With this strategy, tetrahedron tiling is less likely to miss a stable structure.
In order to find a structure that is less stable according to the effective energy model, we adopt an idea that discourages the simulation to explore structures already found.
The idea is similar to the one employed in metadynamics \cite{Parrinello} (However, a collective variable is not required here, since the tetrahedra are already described by a low-dimensional vector and we do not need to recover the potential energy surface. In other words, this method does not suffer from the drawback of metadynamics).
For each stable structure found, positive Gaussian potentials, located at the tetrahedra that compose the structure, are added to the tetrahedron energy landscape $\epsilon (\textbf{r})$ (see Supplemental Material).
When relaunching tetrahedron tiling with the new perturbed tetrahedron energy landscape $\epsilon (\textbf{r})$, these Gaussian potentials artificially destabilize the tetrahedra, thus discouraging the system to visit known structures. Hence the system tends to explore a broader landscape and to build new stable structures.

The other problem is associated with the completeness of the training set.
If the number of structures in the training set is very limited, the known tetrahedra (data points) may be too few to give an effective interpolation, \textit{i.e.}, data points are too far way from each other and important data (tetrahedra) are missing, so that the tetrahedron energy landscape is not fully captured.
We introduce an iterative strategy to systematically improve the training set and the energy landscape.
We note that EDM calculations and tetrahedron tiling simulations can be carried out in an iterative and complementary manner.
While EDM calculations provide energetic input to tetrahedron tiling simulations, new crystal structures predicted by tetrahedron tiling, in return, serve as new samples, which augment the training set.
As the number of loops increases, the accuracy of the tetrahedron energy field gradually improves and is better able to describe the system energetics and ground states.



\begin{figure}
\centering
\includegraphics[width=0.48\textwidth]{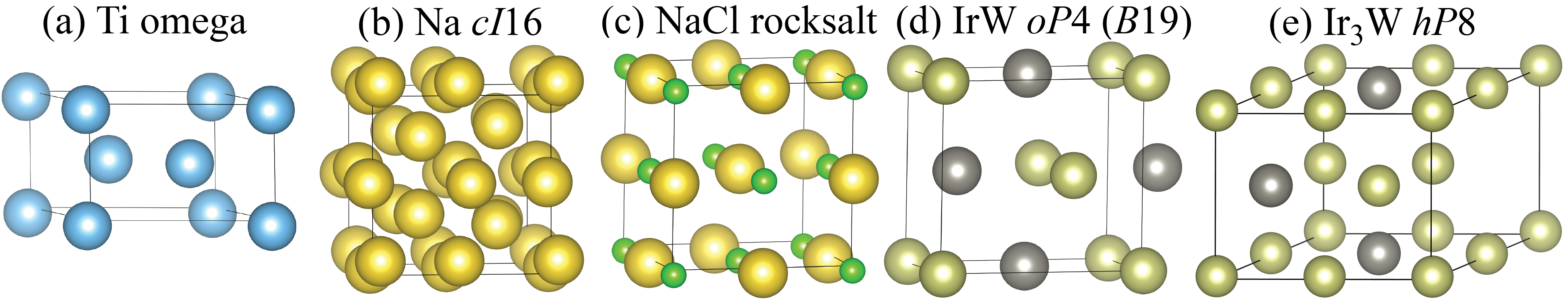}\\
\caption{\label{structs}
Crystal structures predicted by tetrahedron tiling method. (a) Ti omega, (b) Na $cI16$ (a perturbation of bcc phase), (c) NaCl rocksalt (yellow: Na; green: Cl), (d) IrW $oP4$ (yellow: Ir; gray: W. This structure is called $B$19), and (e) Ir$_3$W $hP8$.
These structures are deliberately excluded from training sets.
}
\end{figure}

In order to illustrate the validity of this approach, we provide several examples of application of the method to the crystal structure prediction of titanium, sodium (at 120 GPa), sodium chloride and iridium-tungsten alloys.
Titanium, a simple metal with three major allotropes, allows us to quickly test out the new idea, while sodium has a wide variety of interesting complex crystal structures at high pressure \cite{Gregoryanz08} and is thus representative of challenging CSP problems one may encounter in practice.
Sodium chloride, a typical ionic compound, serves as a prototype of multicomponent system with long-range interaction,
and the success in this material suggests the method's potential in this respect.
Finally we apply the method to iridium-tungsten (Ir-W) alloys, which combine elements that each favors a different lattice: Ir is face-centered cubic (fcc) and W is body-centered cubic (bcc).
This test case most reflects practical settings where the structures of interest are multicomponent ordered alloys. 
We summarize in Table \ref{summary} and Fig. \ref{structs} the crystal structures predicted in these systems.
We note that these structures are deliberately excluded from training sets, and finding them through the tetrahedron tiling method most directly demonstrates the method's ``out-of-sample" predictive ability.

\begin{table}
\caption{A summary of applications of the tetrahedron tiling method \label{summary}}
\begin{tabular}{ccc}
\hline
Materials & Training set &Structures predicted \\ \hline
Ti & fcc, bcc & $hP3$(omega) \\
Na & fcc, bcc, $oP8$ & $cI16$ \\
NaCl & $cP2$(CsCl) & $cF8$(rocksalt) \\
\multirow{2}{*}{Ir-W} & bcc, fcc for  & \multirow{2}{*}{IrW: $oP4$ ($B$19) Ir$_3$W: $hP8$} \\
& Ir, W, and IrW &  \\ \hline
\end{tabular}
\end{table}

We first perform DFT-EDM calculations on a training set and construct a tetrahedron energy landscape for each material.
The DFT calculations are performed with the Vienna Ab initio Simulation Package (VASP) code \cite{VASP} implementing the projector-augmented-wave (PAW) \cite{BLOCHL94} method, with the generalized gradient approximation (GGA) for the exchange-correlation energy, in the form know as Perdew-Burke-Ernzerhof (PBE) \cite{PBE96}.
Based on the tetrahedron energy, we then run tetrahedron-tiling simulations to grow crystal nuclei.
As the size of crystal increases, we detect periodicity and extract periodic unit cells, which we further optimize using DFT and run structure analysis.
We sort these structures by energy and include new structures in the training set for the next iteration.
In the end, structure search completes if no more new stable structures are produced (see Supplemental Material and movies).

\begin{figure}
\centering
\includegraphics[width=0.33\textwidth]{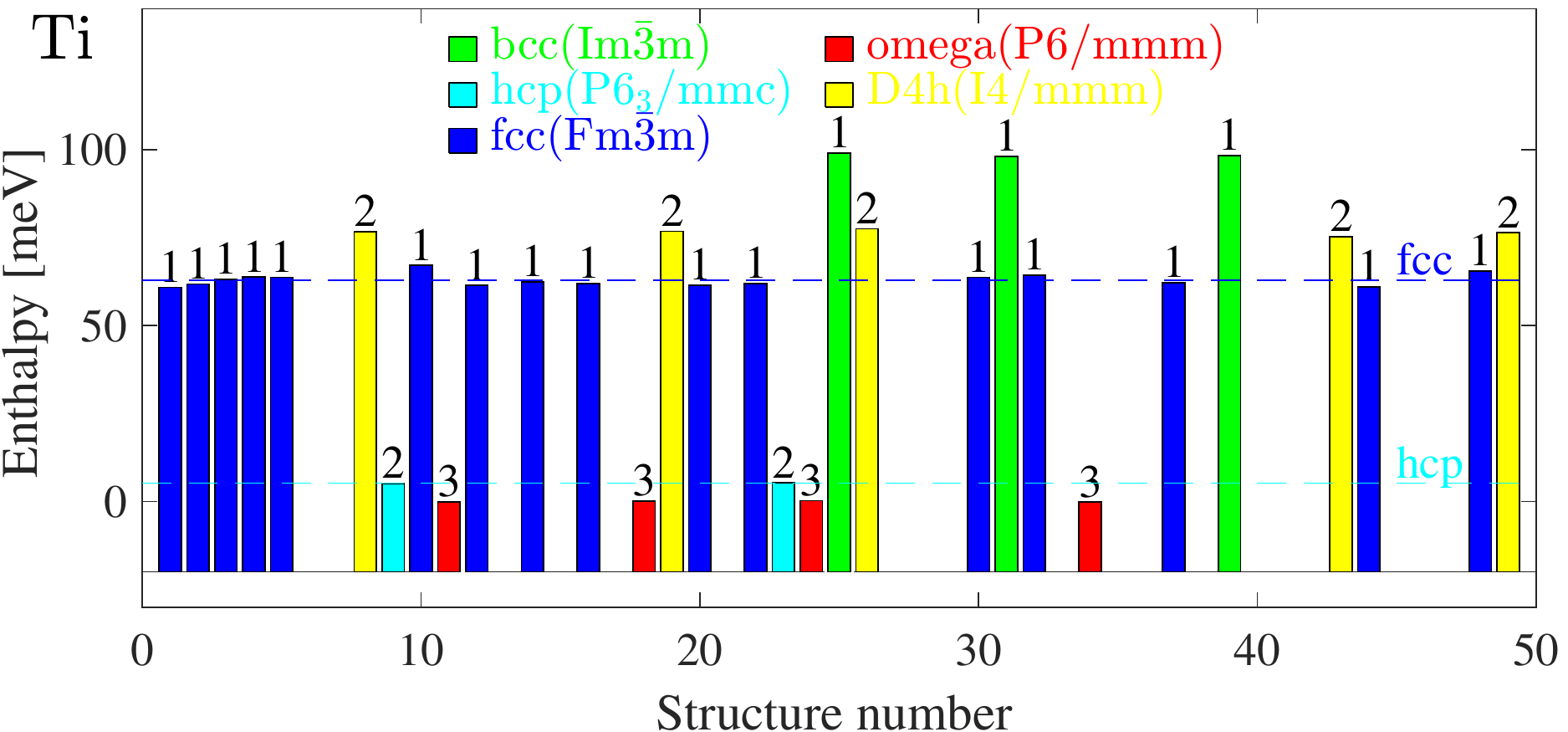}
\includegraphics[width=0.11\textwidth]{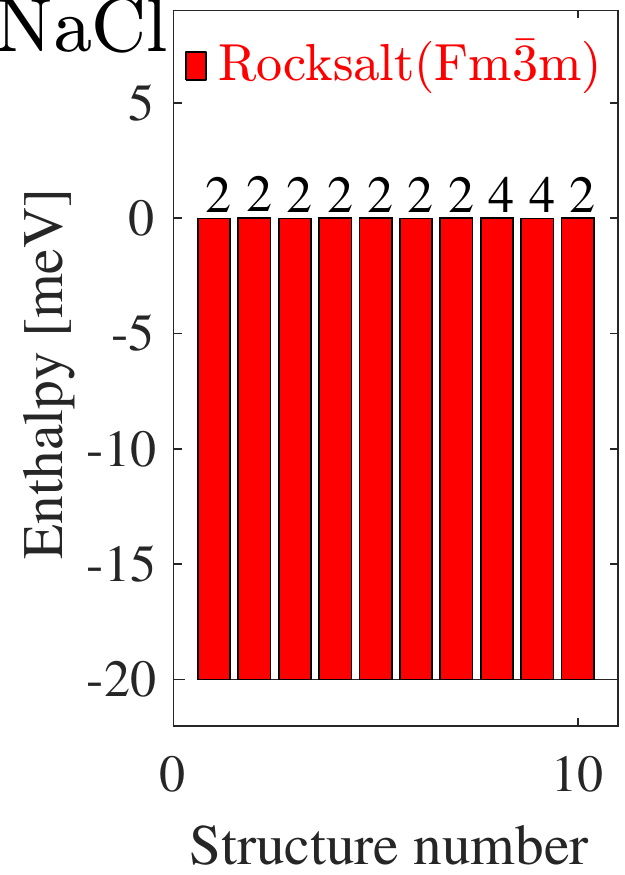}
\includegraphics[width=0.33\textwidth]{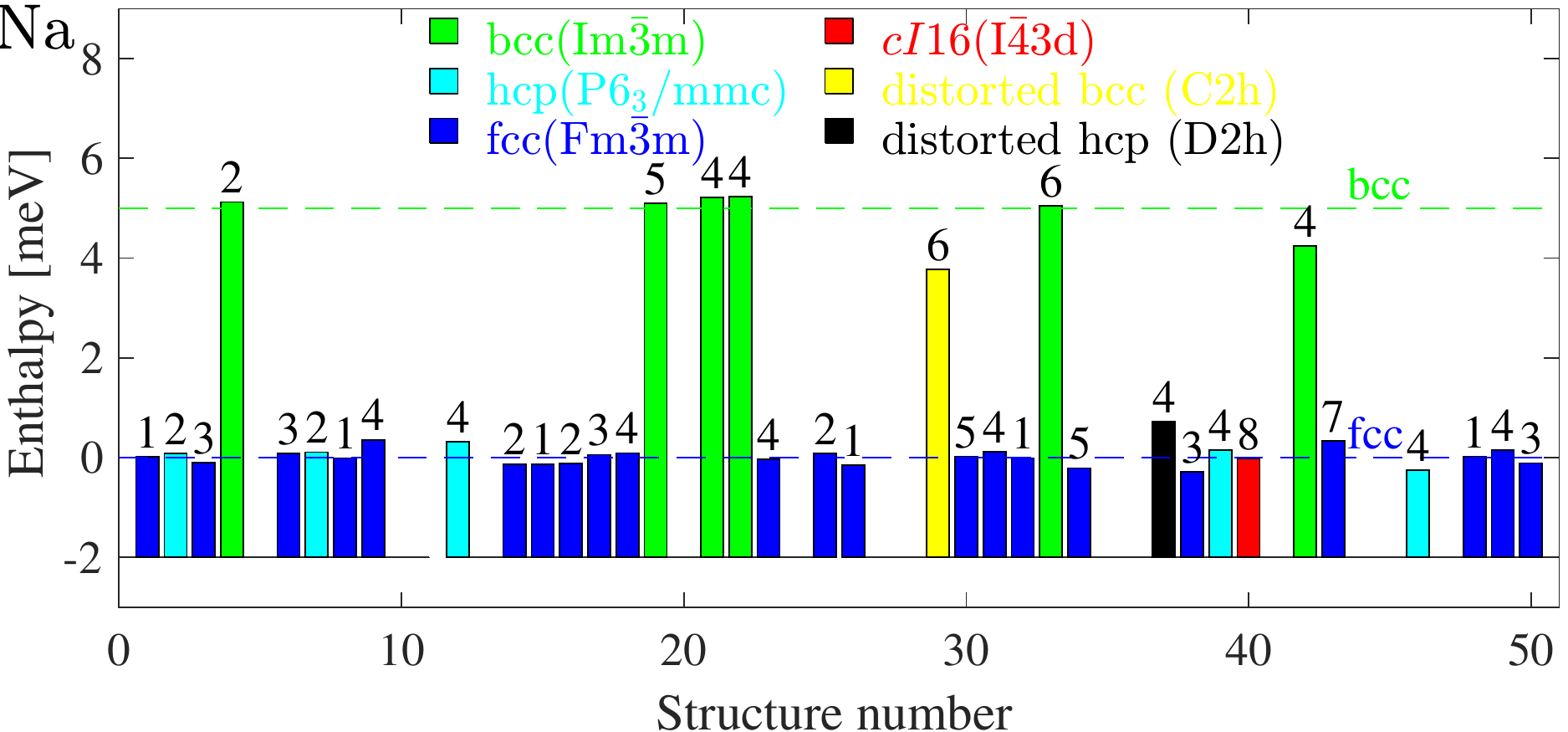}
\includegraphics[width=0.11\textwidth]{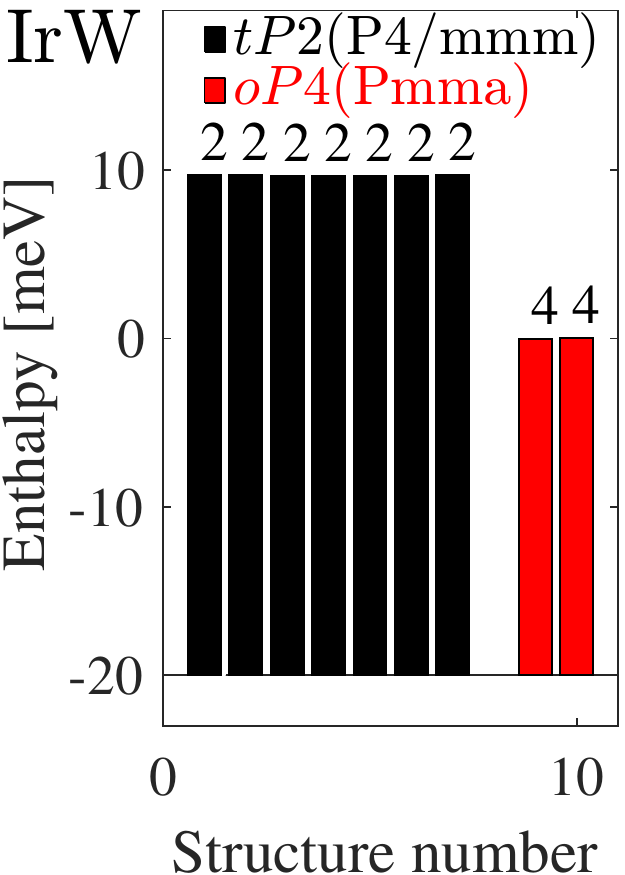}
\caption{\label{screen}
New global minima (red) predicted by the tetrahedron tiling method.
These global minima are deliberately excluded from the training sets, so they are \textit{a priori} unknown.
After tetrahedron tiling simulation builds a structure nucleus (a cluster of atoms, which is usually polymorphic), we extract possible periodic cells (label them with structure number, $x$ axis) and we further optimize them with DFT to nearby local minima.
This figure shows the energetics of the periodic cells after DFT structure optimization, with the number of atoms in each periodic cell labelled above each bar.
The calculations reveal new global minima (red), which demonstrates the method's capability to predict new stable structures.
}
\end{figure}

Tetrahedron-tiling simulations find new crystal structures in three progressive stages. 
Based on the tetrahedron energy function, we first expect tetrahedron tiling simulation to tile and recover, self-consistently, the most stable structure so far in the training set.
Even though this exercise does not constitute a CSP \textit{per se}, it is instructive to illustrate the construction process, and it demonstrates the method's capability to locate a minimum and tile a structure effectively.
In the second stage, we modify the energy landscape and penalize the tetrahedra belonging to the minimum, making it less favorable.
This strategy, destabilizing the minimum found in the previous step, guides the simulation to other minima and iteratively generates a list of stable structures other than the tiled minimum.
This capability allows the simulation to explore and build a diverse range of structures. 
In the final stage, we scrutinizes the predictive power of the approach by examining whether the simulation produces new unknown structures that are even more stable than the current minimum.
Fig. \ref{screen} clearly confirms this capability, as the simulation predicts new global minima (shown in red), which were deliberately excluded from the training sets.

We note that the tetrahedron tiling scheme requires little computational cost and thus is highly efficient, since expensive DFT-EDM calculations are carried out on a very limited number of structures either initially in the training set or found by tetrahedron tiling, and the rest of the energy field is approximated through interpolation.
While the simulation of tetrahedron tiling demands a huge amount of energy evaluation on structures, these calculations are based on the empirical tetrahedron energy landscape, hence having a negligibly small cost.
Indeed, the runtime of our algorithm is competitive compared to other CSP methods.
Though accurate runtime will depend on specific material to study, we here provide a general estimate of computer cost.
A tetrahedron tiling simulation typically completes in $\sim$10--100 h on a single core. The simulation usually generates 10--100 periodic structures which require DFT optimization. 
We repeat this iteration (tetrahedron tiling simulation followed by DFT structure optimization) until the global minima are captured.
In the systems we studied, the global minima are usually discovered within 3 iterations. 

In the case of multicomponent systems, it can be fruitful to explore various possible ordering of the atomic species on the candidate structures found via the tetrahedron tiling method using methods specifically designed for handling ordering problems on a known lattice, such as cluster expansion methods \cite{Sanchez84,vandeWalle02}. In this combination, the two methods are perfectly complementary: tetrahedron tiling can predict the lattice geometry that the cluster expansion is unable to autonomously find while the cluster expansion can handle long-range interactions not accounted for in tetrahedron tiling. (In fact, we have tried this approach for the Ir-W system using the ATAT software \cite{vandeWalle02,ATAT,ATAT2} and the $hP8$ structure was rapidly confirmed as a ground state on the hcp lattice after about 30 small-cell DFT calculations.)

To summarize, we propose a tetrahedron tiling method as an alternative approach to current CSP methods.
This approach employs tetrahedra to combine locally favorable structures into a crystal, a simple and natural way that outperforms slicing crossover.
The method is highly cost-effective, with a low requirement of DFT calculations.
We apply the approach to the crystal structure prediction of titanium, sodium (at 120 GPa), sodium chloride and iridium-tungsten alloys.
It is demonstrated that the approach is capable of (1) finding the global minimum with non-obvious ground states,
(2) generating a series of stable structures with proper perturbation, and (3) predicting new structures outside the training set.

\section*{Acknowledgement}
We thank Drs. Min Yu and Dallas Trinkle for providing EDM code for DFT-EDM calculations. 
This research was supported by NSF under grant DMR-1154895, by ONR under grants N00014-14-1-0055 and N00014-17-1-2202, and by Brown University through the use of the facilities at its Center for Computation and Visualization.
This work uses the Extreme Science and Engineering Discovery Environment (XSEDE), which is supported by National
Science Foundation grant number ACI-1053575.

\end{document}